\documentclass[%
reprint,
 amsmath,amssymb,
 prl,
]{revtex4-1}

\usepackage{graphicx}% Include figure files
\usepackage{dcolumn}% Align table columns on decimal point
\usepackage{bm}% bold math
\renewcommand{\d}{{\rm d}}
\newcommand{\e}{{\rm e}}

\newcommand{\FD}[2]{\frac{\d #1}{\d #2}}
\graphicspath{{Images/}}
\usepackage{todonotes}

\DeclareSymbolFont{AMSb}{U}{msb}{m}{n}
\DeclareMathSymbol{\NSet}{\mathalpha}{AMSb}{"4E}
\DeclareMathSymbol{\ZSet}{\mathalpha}{AMSb}{"5A}
\DeclareMathSymbol{\RSet}{\mathalpha}{AMSb}{"52}
\DeclareMathSymbol{\CSet}{\mathalpha}{AMSb}{"43}

\usepackage{natbib}
\begin{document}

%\preprint{APS/123-QED}

\title{Phase transformation and synchrony for a network of coupled Izhikevich neurons}

\author{\'{A}ine Byrne}
%\email{aine.byrne@ucd.ie}
\affiliation{
School of Mathematics and Statistics, University College Dublin, Belfield, Dublin 4, Ireland.
}%

\date{\today}% It is always \today, today,

\begin{abstract}
A number of recent articles have employed the Lorentz ansatz to reduce a network of Izhikevich neurons to a tractable mean-field description. In this letter, we construct an equivalent phase model for the Izhikevich model and apply the Ott-Antonsen ansatz, to derive the mean field dynamics in terms of the Kuramoto order parameter. In addition, we show that by defining an appropriate order parameter in the voltage-firing rate framework, the conformal mapping of Montbri\'o et al. \cite{Montbrio2015}, which relates the two mean-field descriptions, remains valid. 
\end{abstract}

\pacs{Valid PACS appear here}% PACS, the Physics and Astronomy
\maketitle

\section{\label{sec:intro}Introduction}
Given the immense complexity of the brain, it is only natural that theoreticians seek reduced simplified models of neural dynamics. A recent breakthrough in the field of self-organised systems \cite{Ott2008}, allowed for the derivation of an exact mean-field description for a population of microscopic spiking neurons. Unfortunately, the reduction is only valid for a specific type of neuron model, namely phase oscillator models with harmonic coupling functions, such as the Kuramoto model and the theta-neuron model \cite{Luke2013}. Equivalently, one can apply the Lorentzian ansatz to a network of quadratic integrate-and-fire (QIF) neurons to arrive at a mean-field model which tracks biophysically relevant mesoscopic variables, the firing rate and mean membrane potential \cite{Montbrio2015}.

The theta-neuron reduction was later extended to include a biologically realistic form of synaptic coupling \cite{Coombes2018} and gap junction coupling \cite{Byrne2020}. This mean-field model, dubbed the \emph{next generation neural mass model}, takes a similar form to Wilson-Cowan-like neural mass models. However, the firing rate is now a derived quantity which depends on a dynamic synchrony variable.
A key success of this model is its ability to track the within population synchrony at the mesoscopic level, providing an explicit relationship between measurable mesoscopic variables, such as the firing rate, and the level of synchronisation in the underlying microscopic network. 

More recently, the Lorentzian ansatz has been applied to a network of Izhikevich neurons \cite{Chen2022, Gast2023, Guerreiro2022}. Although the reduction is no longer exact, the reduced mean-field model has been shown to accurately track the population-level activity, namely the firing rate and mean membrane potential. This reduction does not, however, provide a link to the Kuromoto order parameter, and as such, the level of synchrony of the underlying spiking network. 

In this letter we first derive an equivalent phase model for the Izhikevich neuron model. We then apply to Ott-Antonsen ansatz to arrive at a reduced model, which describes the dynamics of the Kuramoto order parameter. Although the reduction is only exact when the adaptation variable is assumed constant or slowly varying, an accurate estimate of the mean-field dynamics can be obtained by coupling the Kuramoto order parameter dynamics to the population-averaged level of adaptation, which is non-constant.  We also challenge the assumption that high firing rates indicate high synchrony and show instead that high amplitude oscillations in the firing rate are a signature of highly synchronised neural activity and suggest that EEG and MEG spectral power could be used to infer the synchrony level of underlying neural populations.

\section{Phase description for the Izhikevich model}
The Izhikevich neuron model is a simplified two-dimensional description of neural dynamics which tracks the neuron's membrane potential $v$ and a phenomenological recovery variable $u$ \cite{Izhikevich2003}. When the membrane potential reaches a predefined threshold $v_{th}$ it is said to have \emph{spiked} and the voltage is reset to a reset value $v_{reset}$. When the neuron spikes the recovery variable is also increased by a set amount $u_{jump}$ to account for the recovery time post spiking. The dynamics of the Izhikevich neuron model are as follows
\begin{align}
C_m\FD{v}{t} &= k(v-v_\theta)(v-v_r) - u + \eta , \label{eq:Izhikevich_v} \\
\FD{u}{t} &= a(b(v-v_r)-u) \label{eq:Izhikevich_u}
\end{align}
with the condition $v(t)\rightarrow v_{reset}$ and $u(t)\rightarrow u(t)+u_{jump}$ when $v(t)>v_{th}$.
Here, $C_m$ is the membrane capacitance, $k$ is the leakage parameter, $v_\theta$ and $v_r$ are the threshold and resting potentials, respectively, $\eta$ is the background drive, $a$ is the recovery time constant, and $b$ relates to the sensitivity of the recovery variable to the subthreshold fluctuations in the membrane potential.

In the absence of adaptation, the Izhikevich neuron model is simply a transformation of the quadratic integrate-and-fire (QIF) model,
\begin{align}
\FD{v}{t} &= v^2+ \eta, \label{eq:QIF} 
\end{align}
with the condition $v(t)\rightarrow v_{reset}$ when $v(t)>v_{th}$.
The QIF model is formally equivalent to the $\theta$-neuron model under the transformation $v =\tan \frac{\theta}{2}$, subject to the conditions $v_{reset}\rightarrow-\infty$ and $v_{th}\rightarrow\infty$. 

The general solution to the QIF model is given as $v_{QIF}(t) =  \frac{\omega}{2} \tan(\frac{1}{2}\omega t+d)$, while the general solution to the Izhikevich neuron model (for fixed $u$) is given as
\begin{align}
v_{izh}(t) = \frac{\omega}{2k} \tan\left(\frac{1}{2}\omega t+d\right) + \frac{v_\theta + v_r}{2} \label{eq:phase_transformation}
\end{align}
where $\omega$ is the frequency of oscillations and $d$ depends on the initial condition. To obtain the correct phase transformation, we note that $$v_{izh}(t|\omega,d) =  \frac{v_{QIF}(t|\omega,d)}{k}+ \frac{v_\theta + v_r}{2}.$$
Given that $v_{QIF} =\tan \frac{\theta}{2}$, we have
$$v_{izh}(\theta)= \frac{1}{k}\tan \frac{\theta}{2}+ \frac{v_\theta + v_r}{2}.$$
Hence, the equivalent phase model for the Izhikevich neuron is given as
\begin{align*}
C_m\FD{\theta}{t} &= 1 - \cos\theta + k\left(1+\cos\theta\right) \left[\eta-u-\frac{k}{4}(v_\theta - v_r)^2\right] \\
\tau_u\FD{u}{t} &= a\left[b\left(\frac{1}{k}\tan \frac{\theta}{2}+\frac{v_\theta + v_r}{2}\right)-u\right],
\end{align*}
where the neuron is said to spike, and $u(t)\rightarrow u(t)+u_{jump}$, when $\theta(t)$ increases through $\pi$.

\section{Ott-Antonsen ansatz}
We now consider a network of $N$ globally coupled conductance-based Izhikevich neurons, with first order synaptic coupling,  
\begin{align}
\tau_s \FD{s}{t} = - s + \frac{\kappa}{N}\sum_{l=1}^N \sum_m \delta(t-t_l^m),\label{eq:Izhikevich_network_s}
\end{align}
where $t_l^m$ is the $m$th spike time of the $l$th neuron and $\kappa$ is the synaptic coupling strength. The results of this study could easily be extended to higher order synaptic coupling. 
Using the phase transformation \eqref{eq:phase_transformation}, we arrive at the following equations
\begin{align}
C_m\FD{\theta_j}{t} &= 1 - \cos\theta_j +k \left(1+\cos\theta_j\right) \left[-\frac{k}{4}(v_\theta - v_r)^2-u_j\right.\nonumber
\\ &\hspace{0.7cm}\left.+ \eta_j +s\left(v_{syn} +\frac{v_\theta + v_r}{2}\right)\right]-s\sin\theta_j,   \label{eq:theta_Izhikevich_network}\\
\FD{u_j}{t} &= a\left[b\left(\frac{1}{k}\tan \frac{\theta_j}{2}+\frac{v_\theta + v_r}{2}\right)-u_j\right],\label{eq:u_Izhikevich_network}
\end{align}
for $j=1,\ldots, N$, and $u_j(t)\rightarrow u_j(t)+u_{jump}$ when $\theta_j(t)$ increases through $\pi$. 

To derive a low-dimensional description of the network we make use of the Ott-Antonsen ansatz \cite{Ott2008}, which can be applied to networks of phase oscillators whose individual dynamics take the following form
\begin{align}
\FD{\theta}{t} = f \e^{i\theta} + h + \bar{f} \e^{i\theta}. \label{eq:general_form_phase}
\end{align}
For our Izhikevich network we have
\begin{align*}
f = \frac{1}{2C_m}(-1 + \Pi+iks ),\hspace{1cm}
h = \frac{1}{C_m}(1 + \Pi),
\end{align*}
where $\Pi=k\left(-\frac{k}{4}(v_\theta - v_r)^2-u+\eta+s\left(v_{syn} +\frac{v_\theta + v_r}{2}\right)\right)$.
Assuming the background drives are chosen from a Lorentzian distribution, with centre $\eta_0$ and full width at half maximum $\Delta$, and employing the Ott-Antonsen ansatz
allows us define the dynamics of the Kuramoto order parameter
\begin{align}
C_m&\FD{Z}{t} = - i\frac{\left(Z - 1\right)^2}{2} + k\frac{\left(Z + 1\right)^2}{2}\left( i\left[-\frac{k}{4}(v_\theta - v_r)^2  \right. \right. \nonumber\\
&\left.\left. - u+ \eta_0 +s\left(v_{syn} +\frac{v_\theta + v_r}{2}\right)\right]-\Delta\right) - s\frac{Z^2-1}{2},\label{eq:Kuramoto_Izhikevich}
\end{align}
where $u = \frac{1}{N}\sum_j u_j$,
\begin{align}
\FD{u}{t} &= a\left(b(V(Z)-v_r)-u\right) + u_{\rm jump} f(Z),\label{eq:MF_u_theta}\\
\tau_s \FD{s}{t} &= - s + \kappa f(Z).\label{eq:MF_s_theta}
\end{align}
The functions $f(Z)$ and $V(Z)$ are given as
\begin{align*}
f(Z) &= \frac{1}{\pi C_m} \frac{1-\left|Z\right|^2}{1+Z+\bar{Z}+\left|Z\right|^2}, \\
V(Z) &=\frac{1}{k} \frac{\text{Im}(Z)}{1+Z+\bar{Z}+\left|Z\right|^2} + \frac{v_\theta + v_r}{2}.
\end{align*}

\begin{figure}
\centering
\includegraphics[width=1\linewidth]{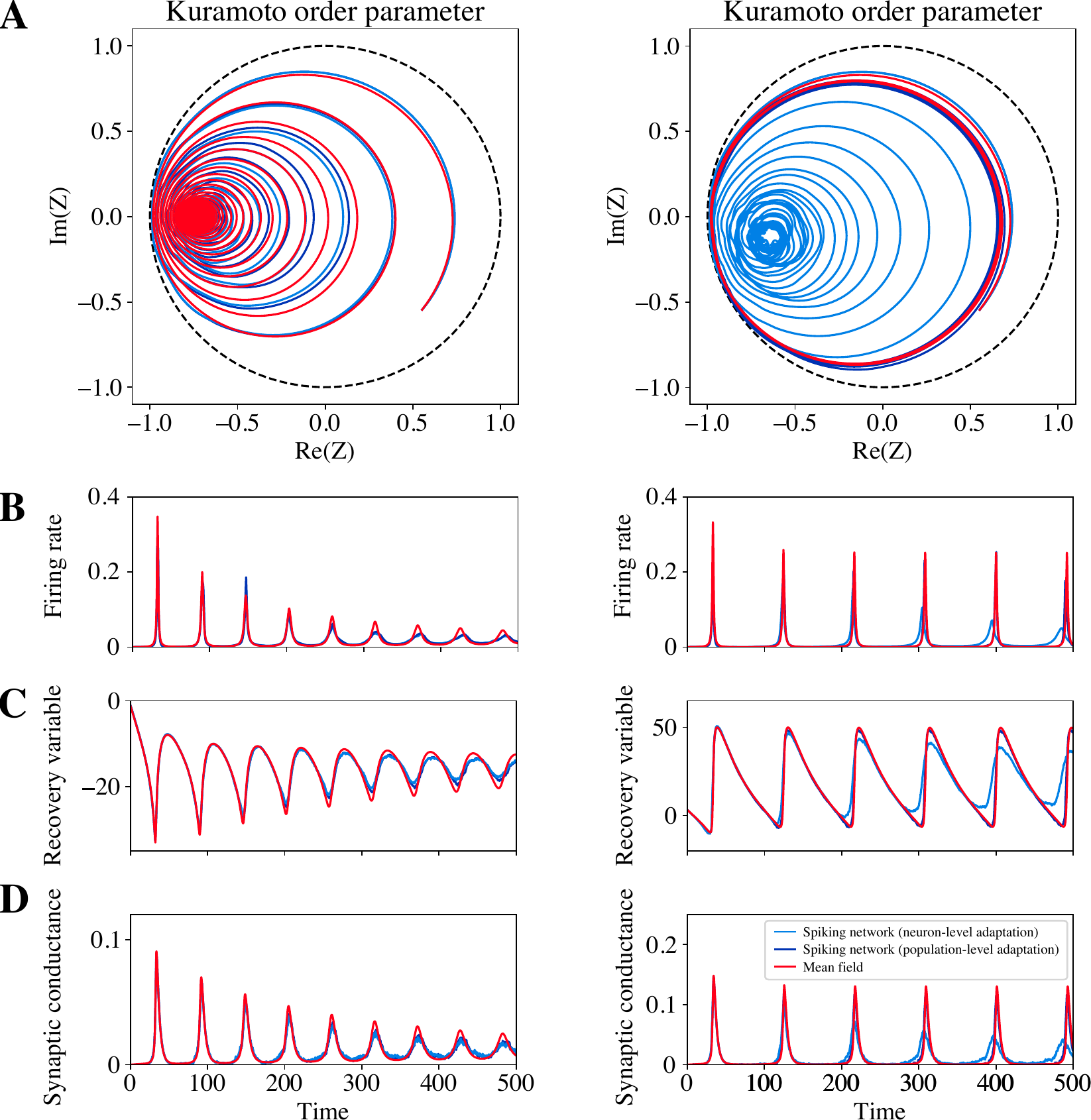}
\caption{\textbf{Mean-field validation:} Comparison of the Kuramoto order parameter (A), firing rate (B), recovery variable (C) and synaptic conductance (D) for the reduced mean field model (red and a simulation of a network of 500 regular spiking Izhikevich neurons with single neuron-level spike-frequency adaptation (light blue) and with population-level spike-frequency adaptation (dark blue).  Left: Weak spike-frequency adaptation ($u_{jump}=10$). Right: Strong spike-frequency adaptation ($u_{jump}=100$). Parameter values: $\eta_0=100$, $\Delta=1$, all other parameters given in the Supplemental Material.}
\label{fig:validity_of_reduction}
\end{figure}

The two most common, and widely studied, neurons in the human cortex are regular spiking (RS) excitatory neurons and fast spiking (FS) inhibitory neurons. For FS neurons, where spike-frequency adaptation is typically not included ($u_{jump}=0$), the mean field equations \eqref{eq:Kuramoto_Izhikevich}--\eqref{eq:MF_s_theta} accurately describe the evolution of the Kuramoto order parameter, the firing rate, the population average of the recovery variable and the average synaptic conductance, respectively, for populations of, as little as, 500 Izhikevich neurons \eqref{eq:Izhikevich_network_s}--\eqref{eq:u_Izhikevich_network} (see Supplemental Material Fig. S1). For RS neurons, however, the accuracy of mean field model decreases as spike-frequency adaptation ($u_{jump}\neq0$) is introduced. This is a result of the mean field description treating the adaptation as a population-level effect, i.e. the adaptation is applied to all neurons in the population rather than just the one that has spiked. If we replace the condition $u_j \rightarrow u_j + u_{jump}$ when $v_j(t)>v_{th}$, with $u_k \rightarrow u_k + u_{jump}/N$ $\forall k$ when $v_j(t)>v_{th}$ in the network simulations, the mean-field equations accurately describe the evolution of the average quantities for network simulations, even for strong spike-frequency adaptation. For weak spike-frequency adaptation ($u_{jump}=10$), the mean field dynamics (red) closely match the spiking network dynamics for both the single neuron-level adaptation (light blue) and the population-level adaptation (dark blue) (Fig. \ref{fig:validity_of_reduction}A). While for strong spike-frequency adaptation ($u_{jump}=100$), the mean field dynamics no longer match the spiking network dynamics for single neuron-level adaptation, but still closely match the dynamics for the network with population-level adaptation (Fig. \ref{fig:validity_of_reduction}B). 
%%%%%%%%%%%%%%%%%%%%%%%%%%%%%%%%%%%%%%%%%%%%%%%%%%%%%%%%%%%%%%%%%%%%%%%%%%%%%%%%%%%%%%%%%%%%%
%%%%%%%%%%%%%%%%%%%%%%%%%%%%%%%%%%%%%%%%%%%%%%%%%%%%%%%%%%%%%%%%%%%%%%%%%%%%%%%%%%%%%%%%%%%%%

%%%%%%%%%%%%%%%%%%%%%%%%%%%%%%%%%%%%%%%%%%%%%%%%%%%%%%%%%%%%%%%%%%%%%%%%%%%%%%%%%%%%%%%%%%%%%
\begin{figure*}
\centering
\includegraphics[width=0.9\linewidth]{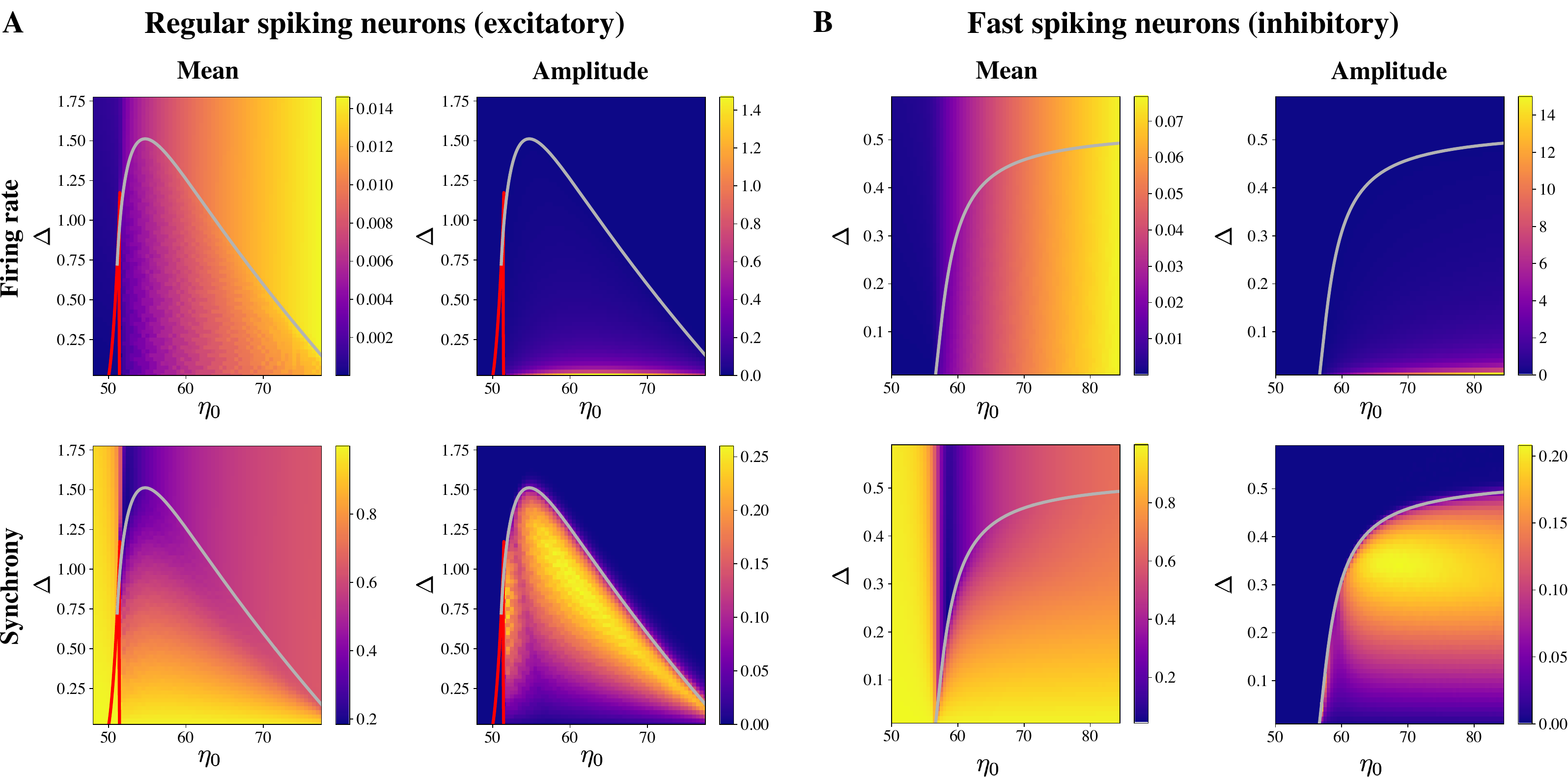}
\caption{\textbf{Isolated populations of regular spiking and fast spiking neurons.} Mean and amplitude of oscillation of the firing rate (top) and within-population synchrony (bottom) as a function of the mean background drive $\eta_0$ and heterogeneity of the drive $\Delta$. Hopf curves are shown in grey and saddle-node curves in red. Both populations exhibit similar behaviours, the mean firing rate increases with $\eta_0$, while the mean synchrony level decreases as $\Delta$ increases. Parameters given in the Supplemental Material.}
\label{fig:single_pop_dynamics}
\end{figure*}
%%%%%%%%%%%%%%%%%%%%%%%%%%%%%%%%%%%%%%%%%%%%%%%%%%%%%%%%%%%%%%%%%%%%%%%%%%%%%%%%%%%%%%%%%%%%%

\section{Mapping to voltage-firing rate representation}
For a network of QIF neurons, there is a direct correspondence between the Kuramoto order parameter and the macroscopic order parameter found by applying the Lorentzian ansatz to the QIF network \cite{Montbrio2015}. This conformal mapping is given as
\begin{align}
Z = \frac{1-\bar{W}}{1-\bar{W}}, \label{eq:conformal_mapping}
\end{align}
where $W = \pi C_m r + iV$, where $r$ is the population firing rate and $V$ is the average membrane potential. The mapping \eqref{eq:conformal_mapping} remains valid for the Izhikevich neuron model. However, the definition of $W$ differs.

As the inflection point for the voltage trace is no longer at $v=0$, the Lorentzian ansatz must be altered to account for the rescaling of $v$ 
\begin{align}
\rho(t,\tilde{v}|\eta) = \frac{1}{\pi}\frac{x(t,\eta)}{\left[\tilde{v}- y(t,\eta)\right]^2 + x^2(t,\eta)}, \label{eq:Lorentzian_ansatz}
\end{align}
where $\tilde{v} = kv-\frac{1}{2}k(v_\theta+v_r) $. The macroscopic order parameter is then defined as
\begin{align*}
W = x + i y = C_m \pi r +  ik\left[v-\frac{1}{2}(v_\theta+v_r) \right].
\end{align*}

%%%%%%%%%%%%%%%%%%%%%%%%%%%%%%%%%%%%%%%%%%%%%%%%%%%%%%%%%%%%%%%%%%%%%%%%%%%%%%%%%%%%%%%%%%%%%
\begin{figure*}
\centering
\includegraphics[width=0.9\linewidth]{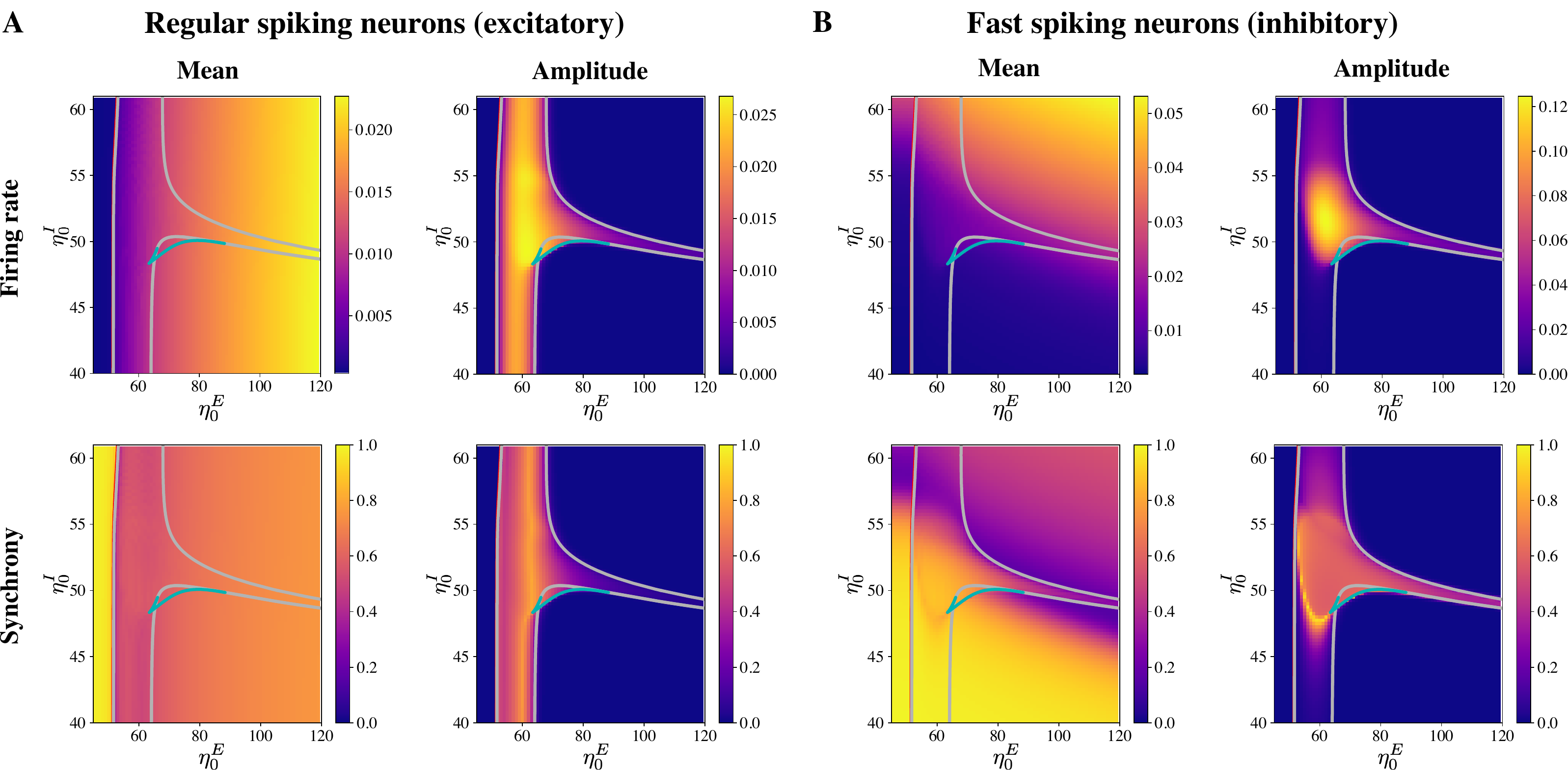}
\caption{\textbf{Network of coupled regular spiking and fast spiking neurons.} Mean and amplitude of oscillation of the firing rate (top) and within-population synchrony (bottom) as a function of the background drives $\eta_0^E$ and $\eta_0^I$. Hopf curves are shown in grey, saddle-node curves in red and the saddle node of periodic orbits in cyan. The oscillation amplitude is maximal when the fast spiking population is highly synchronised with a non-zero firing rate. Parameter values: $\Delta_E=\Delta_I=1$, all other parameters given in the Supplemental Material.}
\label{fig:coupled_pop_dynamics}
\end{figure*}
%%%%%%%%%%%%%%%%%%%%%%%%%%%%%%%%%%%%%%%%%%%%%%%%%%%%%%%%%%%%%%%%%%%%%%%%%%%%%%%%%%%%%%%%%%%%%

Using the modified Lorentzian ansatz \eqref{eq:Lorentzian_ansatz} together with the continuity equation we obtain the following dynamics for $W$
\begin{align*}
\FD{W}{t} &= -iW^2 - \Delta + ik\left[ \eta_0 + I_{syn} - u-\frac{k}{4}(v_\theta - v_r)^2\right].
\end{align*}
Although this equation differs for that proposed by Chen and Campbell \cite{Chen2022}, Gast \emph{et al.} \cite{Gast2023} and Guerreiro \emph{et al.} \cite{Guerreiro2022}, the alternative definitions for $W$ mean that we arrive at the same equations for $r$ and $v$,
\begin{align*}
C_m\FD{r}{t} &= 2krv - k(v_\theta+v_r)r +\frac{k\Delta}{C_m\pi} \\
C_m\FD{v}{t} &= k(v-v_\theta)(v-v_r)+\eta_0 + I_{syn} - u - \frac{(C_m\pi)^2}{k}r^2 , 
\end{align*}
with 
\begin{align*}
\FD{u}{t} &= a\left(b(v-v_r)-u\right) + u_{\rm jump} r,\\
\tau_s \FD{s}{t} &= - s + \kappa r
\end{align*}

\section{\label{sec:firing_rate_synchrony}Relationship between firing rate and synchrony}
A common misconception for mean-field models of neural activity is that a high firing rate implies high synchrony. Using the conformal mapping between the Kuramoto order parameter and the macroscopic firing rate and mean voltage variables, we find that high amplitude oscillations in the firing rate are instead a signature of high synchrony (Fig. \ref{fig:single_pop_dynamics}). We also see high synchrony when the firing rate is low, i.e. when the majority of the neurons are synchronised at rest. Increases in synchrony are also seen as the mean firing rate increases. However, maximal synchrony is seen for parameter values that result in high amplitude oscillations in the firing rate. As expected, increasing the mean background drive ($\eta_0$) increases the firing rate and increasing the level of heterogeneity ($\Delta$) decreases the overall level of synchrony. Interestingly, the level of heterogeneity has little affect on the mean firing rate and other than the abrupt decrease in synchrony where the firing rate becomes non-zero, the mean background drive has little affect on the synchrony. These relationships hold true for isolated populations of both excitatory RS neurons (panel A) and inhibitory FS neurons (panel B). 

Also shown in Fig. \ref{fig:single_pop_dynamics} are the bifurcation curves. The grey curve is a Hopf curve, where oscillations are observed for parameter values under the curve. The red curve in panel A is a saddle node curve, with three fixed points existing between the two lines. Although the saddle node bifurcation is not present for the FS neuron parameter set, the ghost of a saddle node bifurcation can be seen in the sharp increase in firing rate and decrease in synchrony at $\eta_0\approx 55$.

Next we look at coupled populations of RS and FS neurons (see Supplemental Material for equations and parameter values). Again we investigate the relationship between firing rate and synchrony, this time as a function of the mean excitatory background drive ($\eta_0^E$) and the mean inhibitory background drive ($\eta_0^I$) (Fig. \ref{fig:coupled_pop_dynamics}). For the inhibitory FS neurons, there is a clear association between high amplitude oscillations in the firing rate and high synchrony. However, the excitatory RS neurons are moderately to highly synchronised for all values of $\eta_0^E$ and $\eta_0^I$, making it difficult to determine a relationship between firing rate and synchrony in this case. For the RS neurons, changes in $\eta_0^I$ have little to no affect on the mean firing rate and synchrony. Whereas, increase to $\eta_0^E$ and $\eta_0^I$ lead to increases in the mean firing rate for the FS neurons. The amplitude of oscillation is maximal when the FS population is highly synchronised and the firing rate of the FS is non-zero, suggesting that inhibitory neurons plays a pivotal role in generating oscillations. As in Fig. \ref{fig:single_pop_dynamics}, the bifurcation diagram is superimposed on the colormaps. The grey curves represents the Hopf bifurcations and the cyan curve is a saddle-node of periodic orbits bifurcation. There is also a set of saddle-node curves (red curves) close to the left-most Hopf curve, but the window of bi-stability is very small and the two curves are indistinguishable.

\section{\label{sec:discussion}Discussion}

In this work, we derive an equivalent phase model for the Izhikevich neuron model and employ the Ott-Antonsen ansatz to arrive at a next generation neural mass model similar in form to the original model of Byrne and Coombes \citep{Coombes2018}. We show that the conformal mapping of Montbri\'{o} \emph{et al.} remains valid, albeit with a modified order parameter for the voltage-firing rate framework. This mapping provides the explicit relationship between neuronal synchrony and firing rate and average voltage, facilitating the use of such models in whole brain EEG and MEG studies to infer the underlying synchronisation from the observed electrical or magnetic activity. Neuronal synchronisation is a fundamental component of healthy brain function, but too much or too little synchrony can lead to dysfunction \cite{Uhlhaas2009,Mathalon2015,Jiruska2013}. For example, epileptic seizures and motor symptoms in Parkinson's disease are characterised by excessive synchronisation. The ability of next generation neural mass models to track within population synchrony at the mesoscopic level, presents an opportunity to interrogate this delicate balance in neuronal synchrony and investigate the root cause of neurological diseases, such as epilepsy and Parkinson's disease. 

We highlight the misconception that high firing is a signature of synchronised neuronal activity, showing instead a correlation between high amplitude oscillations and high synchrony. Such high amplitude oscillations would manifest as high spectral power in EEG or MEG recordings. Hence, we postulate that EEG and MEG spectral power could be used to infer the level of neural synchronisation.

The original next generation neural mass model was derived from a population of quadratic integrate-and-fire neurons, which have a limited set of firing patterns. The Izhikevich neuron model provides a diverse set of firing responses, associated with different neuron types, enhancing the biological realism of the model. Here, we considered populations of regular spiking and fast spiking neurons, and already we find an array of different bifurcations and behaviour.
We expect an even richer repertoire of dynamics for different neuron types and for networks of multiple different neurons types.

\end{document}